\documentstyle[preprint,eqsecnum,aps]{revtex}
\newcommand{\be}{\begin{equation}}
\newcommand{\ee}{\end{equation}}
\newcommand{\bea}{\begin{eqnarray}}
\newcommand{\eea}{\end{eqnarray}}

\begin{document}
\draft
\title{\hbox{\hspace{20em} \normalsize  PUPT-1477, June 1994} \Large
Renormalization Group Derivation of the Localization Length Exponent in the
Integer Quantum Hall Effect} 
\author{ L. Moriconi}
\address{ Physics Department, Princeton University
\\Jadwin Hall, Princeton, NJ 08544, U.S.A.}
\maketitle
\begin{abstract}
We compute, neglecting possible effects of subleading irrelevant couplings,
the localization length exponent in the
integer quantum Hall effect, for the case of white noise random potentials.
The result obtained is $\nu=2$ for all Landau levels. Our approach consists in
a renormalization group transformation of Landau orbitals, which iterates
the generating functional of Green's functions for the localization problem.
The value of $\nu$ is obtained from the asymptotic form of the renormalization
group mapping. The basic assumptions in our derivation are the existence of a 
scaling law for the localization length and the absence of Landau level mixing.
\end{abstract}

\newpage
\section{Introduction}
The observation of the quantum Hall effect \cite{kep} was a surprise
in the context
of the physics of disordered systems. It is puzzling that there must be, from 
the clear experimental evidence of the Hall conductivity plateaus, delocalized
states which transport current, while it is well-known that in disordered
two-dimensional systems (in the absence of magnetic fields) all states are
localized \cite{aalr}. The picture which has emerged over the last decade
is that near
the center of each Landau level, broadened by the presence of disorder, there is
a specific value of energy, $E_c$, such that states with energy $E$, close
enough to $E_c$, are characterized by a divergent localization length, given by
$\xi(E-E_c) \sim |E-E_c|^{-\nu}$.
This result has been supported by analytical, numerical, 
and experimental studies \cite{hikami,milsok,gurv,aa,chaco,huck,huo,kiv,liu,koch}
. Aoki and Ando \cite{aa} have found numerically
$\nu \stackrel{<}{_\sim} 2$ and $\nu \stackrel{<}{_\sim} 4$, in the case of
$\delta$-impurities, for the first and second Landau levels, respectively.
Subsequent work by many authors confirmed their computation, in the first Landau
level, in particular the semiclassical analysis of Mil'nikov and Sokolov
\cite{milsok}
giving $\nu = 7/3$ and the perturbation approach of Hikami \cite{hikami},
which yields
$\nu = 1.9 \pm 0.2$ (it is worthwhile mentioning, however, that a
careful interpretation of Hikami's result, in the light of the multifractal
structure of wavefunctions, gives $\nu = 2.4 \pm 0.3$, as it was observed in
\cite{jan}). Experimentally \cite{koch}, the value of $\nu$ was
found to be close to $2.3$.
A difficulty in the subject is that, experimentally, Coulomb effects are
expected to modify the result obtained when two-body interactions are neglected
and, therefore, the good agreement between numerical \cite{huck,huo,kiv,liu}
and experimental results
is not very well understood. Another question is the precise dependence of $\nu$
with the Landau level index. A recent numerical analysis \cite{liu}
suggests that for
finite range potentials the value of $\nu$ is universal, being the same for all
Landau levels.

Our aim in this paper is to show a non-perturbative analysis of the problem, 
in the case
of white noise random potentials. We will assume the existence of a scaling
law for $\xi$ and neglect the Landau level mixing, which are the usual
starting points in most of approaches.
The result we found, taking into account only irrelevant couplings of leading
order, is $\nu=2$ for all Landau levels, indicating the
universal behaviour of the localization length exponent in any
transition between plateaus in the integer quantum Hall effect.

The method employed here consists in a redefinition of Landau orbitals, inspired
by the block spin renormalization group \cite{kadanoff}. While in the case of
general Ising models the construction of renormalized Hamiltonians is an
approximation, we will carry an analogous approach in our problem, obtaining
a sequence of exactly renormalized generating functionals of Green's functions.
At each step of the renormalization group iteration, the localization length
decreases by a factor 1/2. Examining, then, the fixed point form of the
consecutive transformations, we are able to obtain the critical exponent
$\nu$.

The plan of the paper is the following. In section II we outline the method,
considering the generating functional of averaged advanced and retarded Green's
functions and showing how a modification of the localization length is
associated to a redefinition of the coupling constants of the system. An
analytical relation for $\nu$ is obtained. In section III we define the
renormalization group transformation and present a detailed computation of
$\nu$. Finally, in section IV, we conclude the analysis and comment on our
results.
      
\section{Outline of the method}

The localization length of a general disordered system may be found through
a computation of $\xi^{-1}(E)$, defined as
\be
\xi^{-1}(E)= -{1\over2}\lim_{  \ |\vec x-\vec x'| \rightarrow \infty}
{{\ln<G_+(\vec x,\vec x';E)G_-(\vec x,\vec x';E)>}
\over {|\vec x-\vec x'|}} \ , \ \label{om1}
\ee
where the brackets in the above expression mean an average over the realizations
of a random potential $V$ and
\be
G_\pm (\vec x,\vec x';E)=<\vec x|(E-H+i0^\pm)^{-1}|\vec x'> \label{om2}
\ee
are the retarded and advanced Green's functions associated to the Hamiltonian
$H=H_0+V$.
A pratical way to study the averaged product of Green's functions is
considering the generating functional (in two spatial dimensions)
\bea
Z[\bar j_1, j_1,\bar j_2,j_2]&=&\int D\bar u_1 Du_1 D\bar u_2 Du_2
\exp \{ i\int d^2x [ \bar u_1(E-H_0+i0^+)u_1 
-\bar u_2(E-H_0+i0^-)u_2   \nonumber \\ 
&+& V(\bar u_1 u_1 - \bar u_2 u_2) +
\bar j_1 \bar u_1 + \bar j_2 \bar u_2 + j_1 u_1 + j_2 u_2 ] \} \ , \ 
\label{om3} 
\eea
where the fields $u_1$ and $u_2$ are chosen to be c-numbers. In this way,
we have
\be
<G_+(\vec x,\vec x';E)G_-(\vec x,\vec x';E)>=\int DV P[V]
{1 \over Z}{{\delta^4 Z} \over
{\delta \bar j_1(\vec x) \delta j_1(\vec x') \delta \bar j_2(\vec x)
\delta j_2(\vec x')}}  \ , \
\label {om4}
\ee
with the derivatives evaluated at $\bar j,j=0$ and $P[V]$ representing the
probability distribution function of the random potential $V$.
We will be concentrated on the gaussian case, given by $<V(\vec x)V(\vec x')>=
\lambda \delta^2(\vec x-\vec x')$.
An interesting trick allows us to write $Z^{-1}[0,0,0,0]$, which appears in
(\ref{om4}), with the same functional form as (\ref{om3}), replacing the
boson fields $u_1$ and $u_2$ by fermionic ones, $v_1$ and $v_2$. Therefore,
in a compact notation, all the information about localization is contained in
the generating functional (we will not write the currents $\bar j$ and $j$ to
avoid cumbersome algebra)
\be
Z=\int DV P[V] D \bar \varphi D \varphi \exp \left \{ i \int d^2 x \left[ \bar \varphi
\left( (E-H_0){\bf A}+i0^+ \right) \varphi + V\bar \varphi {\bf A} \varphi \right] \right\}
\ , \ \label{om5}
\ee
where,
\be
\varphi = {u \choose v} \label{om6}
\ee
\be
u= {u_1 \choose u_2} \ , \ v= {v_1 \choose v_2} \label{om7}
\ee
and the matrix {\bf A}, acting on the components $u$ and $v$ of $\varphi$, is
given by
\be
\pmatrix{\sigma_3 & 0\cr
         0 & \sigma_3\cr}
\ , \ \label{om8}
\ee
where $\sigma_3$ is a Pauli matrix, which acts on the components of $u$ and
$v$. In the case of white noise random potentials, the integration over $V$
in (\ref{om5}) yields
\be
Z=\int D \bar \varphi D \varphi \exp \left \{ i \int d^2 x \left[ \bar \varphi 
\left( (E-H_0){\bf A}+i0^+ \right) \varphi + i\lambda (\bar \varphi {\bf A}
\varphi)^2 \right] \right \}
\ . \ \label{om9}
\ee

In order to study the quantum Hall effect, we may consider the fields $\varphi$
and $\bar \varphi$ as expressed in terms of states belonging only to a specific 
non perturbed Landau level and replace $H_0$ by $E_n=(n+1/2) \hbar \omega_c$,
where $\omega_c$ is the cyclotron frequency. This representation of $\varphi$
corresponds to the assumption of the absence of Landau level mixing. Let us
also suppose that when $E \rightarrow E_n$, the localization length obtained
from (\ref{om9}) has the form $\xi(E-E_n) \sim |E-E_n|^{-\nu}$.

The renormalization group computation that will be performed, is a way to find,
starting from $Z$, a transformed generating functional
\be
\tilde Z=\int D \bar \varphi D \varphi \exp \left \{ i \int d^2 x \left[ \bar \varphi
\left(s(E-E_n,\lambda){\bf A}+i0^+ \right) \varphi + it(E-E_n,\lambda) (\bar \varphi {\bf A}
\varphi)^2 \right] \right \}
\ , \ \label{om10}
\ee 
where $s(E-E_n, \lambda)$ and $t(E-E_n, \lambda)$ are exactly known, with
$s(E-E_n, \lambda) \rightarrow 0$ as $(E-E_n) \rightarrow 0$. We assume now
that the localization length derived from $\tilde Z$ is
$\tilde \xi = {1 \over 2} \xi(E-E_n)$.

Since the number of Grassmann and c-number fields in (\ref{om10}) is the same,
any identical transformation of both fields will produce a unit Jacobian.
In particular, we can redefine the field $\varphi$ by
\be
\varphi \rightarrow \left({\lambda \over t}\right)^{{1 \over 4}}\varphi  \ . \
\label{om11}
\ee
Performing this transformation in $\tilde Z$, we get
\be
\tilde Z=\int D \bar \varphi D \varphi \exp \left \{ i \int d^2 x \left[ \bar
\varphi
\left(s(E-E_n,\lambda)
\left({\lambda \over t}\right)^{{1 \over 2}}
{\bf A}+i0^+ \right) \varphi + i\lambda (\bar 
\varphi{\bf A}
\varphi)^2 \right] \right \}
\ . \ \label{om12}
\ee
Comparing (\ref{om9}) with (\ref{om12}) and using
$\tilde \xi = {1 \over 2} \xi$ we have
\be
\xi(s(E-E_n, \lambda) \left( \lambda / t \right)^{{1 \over 2}})=
{1 \over 2} \xi(E-E_n) \ . \ \label{om13}
\ee
When $E \rightarrow E_n$, we may write $s(E-E_n,\lambda)=s'(0,\lambda)(E-E_n)$.
Substituting the scaling law for $\xi$ in (\ref{om13}) we obtain
\be
\left(s'(0,\lambda)(E-E_n) \left(\lambda / t \right)^{{1 \over 2}}\right)^{-\nu}
= {1 \over 2}(E-E_n)^{-\nu}  \label{om14}
\ee
and, therefore,
\be
\nu = {{\ln2} \over { \ln
\left( s'(0,\lambda) \left(\lambda / t \right)^{{1 \over 2}}\right)}}
\ . \ \label{om15}
\ee
This method for computing $\nu$ is reminiscent from the Kadanoff's block spin
renormalization group \cite{kadanoff}. In fact, as we will see, the similarity to that method
is even deeper than it may be realized from the previous arguments.

\section{Renormalization group analysis}
Let us study the disordered two-dimensional electron gas under the influence of
a magnetic field $B$, as defined on a cylinder of radius $L / {2 \pi}$, where
eventually $L \rightarrow \infty$. In other words, this means that the
wavefunctions of the system must satisfy the periodic boundary condition
$\psi(x,y) = \psi(x+L,y)$. In the absence of disorder, we may use the Landau
gauge, $\vec A = B(-y,0)$, to get the cylinder
eigenfunctions of the $n^{th}$ Landau level,
\be
\psi_{nm}(x,y)=
{1 \over {L^{1/2}}}H_n\left((y-\ell^2mh )/\ell\right)
\exp\left(-(y-\ell^2mh )^2/2\ell^2\right)\exp(imhx)\ , \
\label{rg1}
\ee
where $m=0, \pm 1, ...$, $h \equiv 2 \pi / L$, $\ell = (c/eB)^{1/2}$ is the
magnetic length and
$H_n(x)$ is the Hermite polynomial of order $n$. We can write, thus,
\be
\varphi(x,y)=\left( {h \over {2 \pi} } \right )^{1/2} \sum_m 
H_n\left((y-\ell^2mh)/\ell\right)
\exp\left(-(y-\ell^2mh )^2/2\ell^2\right)\exp(imhx)
a_{mh} \ , \ \label{rg2}
\ee
with $\bar \varphi(x,y)$ analogously defined in terms of $\bar a_{mh}$.
The fact that the above fields $a_{mh}$ and $\bar a_{mh}$ depend on a single
variable $m$, reflects the well-known mapping between the quantum Hall
effect defined on the cylinder and a one-dimensional fermion system.
In particular, the inverse localization length, previously expressed by
(\ref{om1}), may be
rewritten in the one-dimensional formalism as
\be
\xi^{-1}(E)= -{1\over2}\lim_{  \ | k - k'| \rightarrow \infty}
{{\ln<\left ( \bar a_k \ {\bf P_+} a_{k'} \right )
\left ( \bar a_{k'} {\bf P_-} a_k \right )>}
\over {| k - k'|}} \ , \ \label{rg3}
\ee
where the $4\times4$ matrices ${\bf P_+}$ and ${\bf P_-}$,
\be
{\bf P_{\pm}}= {1 \over 2}
\pmatrix{I \pm \sigma_3 & 0\cr
         0 & 0\cr}
\ , \ \label {rg4}
\ee
are the projectors which single out the retarded and advanced c-number
components of $a_k$, respectively.

Using the above expansion of $\varphi(x,y)$, we can express the
generating functional (\ref{om9}) as
\bea
&&Z=\int \prod_{m} D \bar a_{mh} D a_{mh} \exp \left \{ i \sum_{m} \bar a_{mh} \right. 
\left( (E-E_n){\bf A}+i0^+ \right) a_{mh} \nonumber \\
&&\left. - {{\lambda h} \over {2 \pi}} \sum_{m_1,m_2,m_3} g_n((m_1-m_2)h,(m_3-m_2)h) 
\left ( \bar a_{m_1h} {\bf A} a_{m_2h} \right )
( \bar a_{m_3h} {\bf A} a_{(m_1-m_2+m_3)h} ) \right \}
\ , \ \label{rg5}
\eea
where
\bea
&&g_n((m_1-m_2)h,(m_3-m_2)h)= \nonumber \\
&&\int dy H_n\left((y-\ell^2m_1h)/\ell\right)
\exp\left(-(y-\ell^2m_1h )^2/2\ell^2\right)
H_n\left((y-\ell^2 m_2h)/\ell\right) \cdot \nonumber \\
&&\cdot \exp\left(-(y-\ell^2m_2h )^2/2\ell^2\right)
H_n\left((y-\ell^2 m_3h)/\ell\right)
\exp\left(-(y-\ell^2m_3h )^2/2\ell^2\right) \cdot \nonumber \\
&&\cdot H_n\left((y-\ell^2(m_1-m_2+ m_3)h)/\ell\right)
\exp\left(-(y-\ell^2(m_1-m_2+m_3)h )^2/2\ell^2\right) \ . \ \label{rg6}
\eea
The continuum limit, $h \rightarrow 0$, is, as usual, performed by means of
the replacements	
\bea
&&mh \rightarrow k  \\ \label{rg7}
&&\sum_m \rightarrow {1 \over h}  \int dk  \\ \label{rg8}
&&a_{mh} \rightarrow h^{1/2}a_k \ . \ \label{rg9}
\eea
The continuum definition of (\ref{rg5}) is, then,
\bea
&&Z=\int D\bar a_k D a_k \exp \left[ i \int \right. dk \bar a_k \left((E-E_n)
{\bf A}
+ i0^+ \right) a_k  \nonumber \\
&-&\left.  (\lambda / 2 \pi ) \int dk_1 dk_2 dk_3 g_n(k_1-k_2,k_3-k_2)
(\bar a_{k_1} {\bf A} a_{k_2})(\bar a_{k_3} {\bf A} a_{(k_1-k_2+k_3)}) \right]
\ . \ \label{rg10}
\eea
The generating functional (\ref{rg5}) describes the dynamics of a multicomponent
field defined on a one-dimensional discrete lattice. The first step of the
renormalization group transformation consists of the definition of the set
of fields $a_{(+,mh)} \equiv a_{2mh}$ and $a_{(-,mh)} \equiv a_{(2m+1)h}$,
related to the even and odd sites of the lattice, respectively. The point here
is that the localization length derived from $a_{(+,mh)}$ or $a_{(-,mh)}$ is
precisely $1/2$ of the localization length associated to $a_{mh}$, as may be
seen from relation (\ref{rg3}). Expressing now the generating functional
(\ref{rg5}) in terms of $a_{(+,mh)}$ and $a_{(-,mh)}$, we get
\be
Z=\int \prod_m D \bar a_{(+,mh)} D a_{(+,mh)} D \bar a_{(-,mh)} D a_{(-,mh)}
\exp \left \{ i \sum_{m}S^{(2)}_{m} - {{\lambda h}\over {2 \pi}}
\sum_{m_1,m_2,m_3}S^{(4)}_{(m_1,m_2,m_3)} \right \}
\ , \ \label{rg11}
\ee 
with
\be
S^{(2)}_{m} =  \bar a_{(+,mh)} \left ( (E-E_n) {\bf A} +
i0^+ \right )a_{(+,mh)} + \bar a_{(-,mh)} \left ( (E-E_n) {\bf A} +
i0^+ \right )a_{(-,mh)}  \label{rg12}
\ee 
and
\bea
&&S^{(4)}_{(m_1,m_2,m_3)} = 
\bar a_{(+,m_1h)}{\bf A}a_{(+,m_2h)} \left[ g_n( 2(m_1-m_2)h,2(m_3-m_2)h)
\bar a_{(+,m_3h)}{\bf A}a_{(+,(m_1-m_2+m_3)h)} \right. \nonumber \\
&&\left. +g_n( 2(m_1-m_2)h,
2(m_3-m_2)h+h)
\bar a_{(-,m_3h)}{\bf A}a_{(-,(m_1-m_2+m_3)h)} \right]
+ \bar a_{(+,m_1h)}{\bf A}a_{(-,m_2h)} \cdot \nonumber \\
&&\cdot \left[ g_n( 2(m_1-m_2)h-h,2(m_3-m_2)h-h)
\bar a_{(+,m_3h)}{\bf A}a_{(-,(m_1-m_2+m_3-1)h)} \right. \nonumber \\
&&\left. +g_n(2(m_1-m_2)h-h,
2(m_3-m_2)h)
\bar a_{(-,m_3h)}{\bf A}a_{(+,(m_1-m_2+m_3)h)} \right]  \nonumber \\
&&+\bar a_{(-,m_1h)}{\bf A}a_{(+,m_2h)} \left[ g_n(2(m_1-m_2)h+h,2(m_3-m_2)h)
\bar a_{(+,m_3h)}{\bf A}a_{(-,(m_1-m_2+m_3)h)} \right. \nonumber \\
&&\left. + g_n( 2(m_1-m_2)h+h,2(m_3-m_2)h+h) \bar a_{(-,m_3h)}{\bf A}
a_{(+,(m_1-m_2+m_3+1)h)} \right] \nonumber \\
&&+\bar a_{(-,m_1h)}{\bf A}a_{(-,m_2h)} \left[ g_n(2(m_1-m_2)h,2(m_3-m_2)h+h)
\bar a_{(+,m_3h)}{\bf A}a_{(+,(m_1-m_2+m_3)h)} \right.  \nonumber \\
&&\left. g_n( 2(m_1-m_2)h,
2(m_3-m_2)h)
\bar a_{(-,m_3h)}{\bf A}a_{(-,(m_1-m_2+m_3)h)} \right] \ . \ 
\label{rg13} 
\eea

We would like to find, in view of the arguments presented in section II, a
renormalized generating functional, related to anyone of the fields
$a_{(+,mh)}$ or $a_{(-,mh)}$, which has the same structure as (\ref{rg5}).
In order to do that, let us first note
that there are in $S^{(4)}$ some coefficients and fields which differ
from the others only by terms of the order of $h$, when $h \rightarrow 0$.
We may, therefore, perform an expansion of (\ref{rg13}) up to ${\cal O}(h)$.
It is
useful, for this purpose, to define
\bea
g_n^{(1)}(x,y) = {\partial \over {\partial x}} g_n(x,y)  \\ \label{rg14} 
g_n^{(2)}(x,y) = {\partial \over {\partial y}} g_n(x,y)   \label{rg15}
\eea
and also
\bea
\Phi_{mh} = {a_{(+,mh)} \choose a_{(-,mh)}}  \\ \label{rg16} 
\delta \Phi_{mh} = \Phi_{mh} - \Phi_{(m-1)h} \ . \ \label{rg17}
\eea
The result of the expansion gives, after straightforward computations,
\be
Z = \int \prod_m D \bar \Phi_{mh} D \Phi_{mh} \exp \left \{ iS + i \delta S
\right \} \ , \ \label{rg18}
\ee
where
\bea
&&S = \sum_m \bar \Phi_{mh} \left ( (E-E_n) {\bf A} + i0^+ \right ) \Phi_{mh}
+ {{i \lambda h} \over {2 \pi}} \sum_{\{m\}} g_n \cdot
\left [ \left ( \bar \Phi_{m_1h}{\bf A}\Phi_{m_2h} \right) \right. 
\cdot \nonumber \\ 
&&\left. \cdot \left( \Phi_{m_3h}{\bf A} \Phi_{(m_1-m_2+m_3)h} \right)+
\left ( \bar \Phi_{m_1h}{\bf A} \sigma_1 \Phi_{m_2h} \right) \left( \bar
\Phi_{m_3h}{\bf A} \sigma_1 \Phi_{(m_1-m_2+m_3)h} \right) \right]
\label{rg19}
\eea
and
\bea
&&\delta S = i \lambda \sum_{m_1,m_2,m_3}
\left \{ { {h^2} \over {2 \pi}} g_n^{(2)} \cdot
\left ( \bar \Phi_{m_1h}{ {(1+\sigma_3)} \over 2}{\bf A}
\Phi_{m_2h} \right )   
\left( \bar \Phi_{m_3h}{{(1-\sigma_3)} \over 2} {\bf A}
\Phi_{(m_1-m_2+m_3)h} \right) \right. \nonumber \\ 
&&-{ h^2 \over {2 \pi}} \left( g_n^{(1)} +g_n^{(2)} \right)
\cdot \left ( \bar \Phi_{m_1h}{ {(\sigma_1+i\sigma_2)} \over 2}{\bf A}
\Phi_{m_2h} \right )
\left ( \bar \Phi_{m_3h}{ {(\sigma_1+i\sigma_2)} \over 2}
{\bf A}\Phi_{(m_1-m_2+m_3)h} \right ) \nonumber \\
&&-{ {h^2} \over {4 \pi}}g_n^{(1)} \cdot
\left [ \left ( \bar \Phi_{m_1h}{ {(\sigma_1+i\sigma_2)} \over 2}{\bf A}
\Phi_{m_2h} \right ) \left ( \bar \Phi_{m_3h}{ {(\sigma_1-i\sigma_2)} \over 2}
{\bf A}\Phi_{(m_1-m_2+m_3)h} \right ) + (+ \longleftrightarrow -) \right ]
\nonumber \\
&& \left. -{ h \over {2 \pi}}g_n
\cdot \left ( \bar \Phi_{m_1h}{ {(\sigma_1+i\sigma_2)} \over 2}{\bf A}
\Phi_{m_2h} \right ) 
\left ( \bar \Phi_{m_3h}{ {(\sigma_1+i\sigma_2)} \over 2}
{\bf A} \delta \Phi_{(m_1-m_2+m_3)h} \right ) + c.c  \right \} \ . \ 
\label{rg20}
\eea
In (\ref{rg19}) and (\ref{rg20}) the
functions $g_n$, $g_n^{(1)}$ and $g_n^{(2)}$ are implicitly defined on the
point $(x,y) = (2(m_1-m_2)h,2(m_3-m_2)h)$.
In these equations, the Pauli matrices
$\sigma_1$, $\sigma_2$ and $\sigma_3$ act on the components $a_{(+,mh)}$ and
$a_{(-,mh)}$ of $\Phi_{mh}$.

Considering that near the ``critical point", $E=E_n$, the field 
$\Phi_{mh}$ will not fluctuate considerably in a certain range of sites
of the one-dimensional lattice, we may write $\delta \Phi_{mh}
\rightarrow h^{3/2} {\partial \over {\partial k}} \Phi_k$ as $h \rightarrow
0$, and it is immediately seen that $\delta S$ vanishes in the continuum limit.
We will
study, thus, only the action $S$, given by $(\ref{rg19})$, in the next
computations. 
It is interesting to note, however, that there is an $O(2)$
symmetry in the problem, expressing the invariance of $S$ under the
mapping $\Phi_{mh} \rightarrow \exp (i \theta \sigma_1) \Phi_{mh}$,
which is explicitly broken by $\delta S$.

We can use a $SU(2)$ transformation, given by
$U^+ \sigma_1 U = \sigma_3$, $\Phi_{mh} \rightarrow U \Phi_{mh}$ and
$\bar \Phi_{mh} \rightarrow \bar \Phi_{mh} U^+$, to find, from (\ref{rg19}),
$Z=Z_+Z_-$,
where
\bea
&&Z_+=\int \prod_{m} D \bar a_{(+,mh)} D a_{(+,mh)} \exp
\left \{ i \sum_{m} \bar a_{(+,mh)}
\left( (E-E_n){\bf A}+i0^+ \right) a_{(+,mh)}
-2 {{\lambda h} \over {2 \pi}}
\cdot \right. \nonumber \\
&&\left. \cdot \sum_{m_1,m_2,m_3}
g_n(2(m_1-m_2)h,2(m_3-m_2)h)
\left ( \bar a_{(+,m_1h)} {\bf A} a_{(+,m_2h)} \right )
\left( \bar a_{(+,m_3h)} {\bf A} a_{(+,(m_1-m_2+m_3)h)} \right)
\right \} \nonumber \\
\label{rg21}
\eea
and similarly for $Z_-$. We will prove now that the above $SU(2)$ rotation
does not modify the localization length associated to each one of the components
of $\Phi_{mh}$
and so (\ref{rg21}) gives us a theory with a localization length which is $1/2$
of that one derived from $Z$, defined by (\ref{rg5}).
In order to understand it, let us note that, according to (\ref{rg3}), we have to
study how the correlation function
\be
<\left ( \bar a_{(+,mh)}  \ {\bf P_+} a_{(+,m'h)} \right )
\left ( \bar a_{(+,m'h)} {\bf P_-} a_{(+,mh)} \right )> \label{rg22}
\ee
is modified when the $SU(2)$ rotation,
\be
a_{(+,mh)} \rightarrow {1 \over {\sqrt 2}} (a_{(+,mh)}+a_{(-,mh)})
\label{rg23}
\ee
is performed. Therefore, substituting (\ref{rg23}) in (\ref{rg22}), we get
\bea
&&{1 \over 4}
< \left [ \left(\bar a_{(+,mh)}+ \bar a_{(-,mh)} \right) {\bf P_+}
\left ( a_{(+,m'h)}+a_{(-,m'h)} \right) \right ]
\left [ \left (\bar a_{(+,m'h)}+ \bar a_{(-,m'h)} \right) {\bf P_-}
\cdot \right.  \nonumber \\
&&\left. \cdot \left ( a_{(+,mh)}+a_{(-,mh)} \right) \right ] > \ . \ 
\label{rg24}
\eea
The above correlation function must be evaluated in the theory defined by
$Z=Z_+Z_-$.
The factorization of $Z$ implies that (\ref{rg24}) is exactly
\be
{1 \over 2} \left \{ <\left ( \bar a_{(+,mh)}  \ {\bf P_+} a_{(+,m'h)} \right )
\left ( \bar a_{(+,m'h)} {\bf P_-} a_{(+,mh)} \right )>+
|<\left( \bar a_{(+,mh)}  \ {\bf P_+} a_{(+,m'h)} \right )>|^2 \right \} \ , \
\label{rg25}
\ee
where we used, between (\ref{rg24}) and (\ref{rg25}), the fact that the fields
$a_{(+,mh)}$ and $a_{(-,mh)}$ have identical correlation functions. Now,
according to the general result for the averaged Green's functions, stating
that $|<G(k,k';E)>|^2$ vanishes faster than $<|G(k,k';E)|^2>$ when $|k-k'|
\rightarrow \infty$, we can see that the second term in (\ref{rg25}) may be
neglected, in order to compute the localization length. We find, then, that
the only effect of the $SU(2)$ transformation is a factor of $1/2$ multiplying
the original correlation function (\ref{rg22}), which, from (\ref{rg3}),
does not modify the value obtained for $\xi(E)$. 

In the continuum limit, $Z_+$ is written as
\bea
&&Z_+=\int D\bar a_{(+,k)} D a_{(+,k)} \exp
\left[ i \int \right. dk \bar a_{(+,k)} \left((E-E_n)
{\bf A}
+ i0^+ \right) a_{(+,k)}  \nonumber \\
&-&2 {\lambda \over { 2 \pi}} 
\left. \int dk_1 dk_2 dk_3 g_n(2(k_1-k_2),2(k_3-k_2))
\left( \bar a_{(+,k_1)} {\bf A} a_{(+,k_2)} \right) \left( \bar a_{(+,k_3)}
{\bf A}
a_{(+,k_1-k_2+k_3)} \right) \right]
\ . \ \nonumber \\
\label{rg26}
\eea
We can, in the same way, iterate $Z_+$ to find a new generating functional
$Z_{++}$ and so on. After $N$ iterations of $Z$, we will have
\bea
&&Z^{(N)}=\int D\bar a_k D a_k \exp
\left[ i \int \right. dk \bar a_k \left((E-E_n)
{\bf A}
+ i0^+ \right) a_k  \nonumber \\
&-&2^N {\lambda \over { 2 \pi}}
\left. \int dk_1 dk_2 dk_3 g_n(2^N(k_1-k_2),2^N(k_3-k_2))
( \bar a_{k_1} {\bf A} a_{k_2} ) ( \bar a_{k_3}
{\bf A}
a_{(k_1-k_2+k_3)} ) \right]
\ . \ \nonumber \\
\label{rg27}
\eea
One may observe that the above renormalized generating functional can be
obtained directly from $Z$, eq. (\ref{rg10}), through the mapping
$a_k \rightarrow {1 \over {2^{N/2}}} a_{(k/2^N)}$.
Regarding $g_n$ as a distribution (see the appendix A), we find,
from (\ref{rg6}),
\be
\lim_{N \rightarrow \infty} g_n(2^N(k_1-k_2),2^N(k_3-k_2)) = 
{C_n \over {2^{2N}}} \delta(k_1-k_2) \delta(k_3-k_2) \ , \ \label{rg28}
\ee
where it is unimportant to know the exact value of $C_n$. Substituting
(\ref{rg28}) in (\ref{rg27}), we get the asymptotic form of $Z^{(N)}$,
\be
Z^{(N)}=\int D\bar a_k D a_k \exp \left[ i \int dk
\bar a_k \left((E-E_n)
{\bf A}
+ i0^+ \right) a_k -  { {\lambda } \over { 2 \pi}} { {C_n} \over {2^N}}
\int dk (\bar a_k {\bf A} a_k)^2 \right] \ , \ \label{rg29}
\ee
where the term proportional to $\lambda$ has to be considered as a point
splitted product of fields.
The above definition of the asymptotic form of $Z^{(N)}$
is equivalent to neglecting
irrelevant terms of subleading order, in the framework of the renormalization
group transformation we have been studying. More precisely, let us consider a
power expansion of the product of four fields
\bea
&&( \bar a_{k_1} {\bf A} a_{k_2} ) ( \bar a_{k_3}
{\bf A} a_{(k_1-k_2+k_3)} ) =
\sum_{n_1,n_2,n_3=0}^\infty
c(n_1,n_2,n_3)(k_2-k_1)^{n_1}(k_3-k_1)^{n_2}(k_3-k_2)^{n_3}
\cdot \nonumber \\
&&\cdot \left( \bar a_{k_1} {\bf A}( {{\partial^{n_1}} \over
{\partial k_1^{n_1}}} 
a_{k_1}) \right) \left( ( {{\partial^{n_2}} \over
{\partial k_1^{n_2}}}\bar a_{k_1}
){\bf A} ( {{\partial^{n_3}} \over
{\partial k_1^{n_3}}} a_{k_1} ) \right) \ . \ \label{rg30}
\eea
In (\ref{rg30}) there is a sequence of products of operators, defined at the
same point $k_1$. This may be regularized through the point splitting
procedure, which consists, if one wants to analyze 
the generating functional $Z^{(N)}$, of chosing a scale $\tau \ll 1/ 
(2^N \ell)$ and
writing the product of four generic operators $O_1(k)$, $O_2(k)$, $O_3(k)$
and $O_4(k)$ as
\bea
&&O_1(k)O_2(k)O_3(k)O_4(k) = { 1 \over {\tau^2 }} \int^{\tau / 2}_{- \tau /2}
\int^{\tau / 2}_{- \tau /2} dp dq O_1(k)O_2(k+p)O_3(k+p+q)O_4(k+q) 
\ . \ \nonumber \\
\label{rg31}
\eea
The limit $\tau \rightarrow 0$ is performed after the computation physical
quantities.
Substituting (\ref{rg30}) in (\ref{rg27}) we will have a sum of terms 
with coefficients given by
\bea
2^N {\lambda \over {2 \pi}} \int dp dq 
g_n(2^Np,2^Nq)
c(n_1,n_2,n_3)p^{n_1}(p+q)^{n_2}q^{n_3}
\sim {\lambda \over {2 \pi}} {{c(n_1,n_2,n_3)} \over {2^{N(n_1+n_2+n_3+1)}}}
\ . \
\label{rg32}
\eea 
In the language of the renormalization group theory, (\ref{rg32}) means that
we have an infinite set of irrelevant couplings.
The asymptotic form (\ref{rg29})
includes the effects of only the leading term in this expansion, corresponding
to $n_1,n_2,n_3=0$.
We see, therefore, from (\ref{rg29}), that $Z^{(N)} \rightarrow Z^{(N+1)}$ is
obtained from $(E-E_n) \rightarrow (E-E_n)$ and
$\lambda \rightarrow \lambda /2$. According to the discussion of section
II, the functions $s$ and $t$ for this mapping are $s(x,y)=x$ and
$t(x,y)=y/2$. Applying now eq. (\ref{om15}) we get $\nu=2$.

\section{Conclusion}
Using the two basic assumptions of the existence of a localization length
scaling law and the absence of Landau level mixing, we were able to find,
neglecting irrelevant couplings of subleading order, the
value $\nu=2$ for the localization length exponent,
in all Landau levels.
The method of computation is based on an exact sequence of renormalization group
transformations. The fundamental ingredient in the analysis is that after
each transformation of the generating functional of Green's functions,
we obtain a theory which gives $1/2$ of the previous localization length.

It is possible that the neglected terms (the one containing $\delta \Phi_{mh}$
in (\ref{rg20}) and the terms with coefficients expressed by (\ref{rg32}), for
$n_1+n_2+n_3 \geq 1$) have a participation in the evaluation
of $\nu$, similar to what happens when dangerous irrelevant variables come into
play in the statistical mechanics of phase transitions.
The subleading terms could account for the fact that there is
already some numerical
evidence establishing that $\nu \sim 2.3$ \cite{huck}.
Related to it, we advance that $\nu=2$ may be associated to a possible 
intermediate scaling region in the
crossover between the localization length exponent obtained via classical
percolation theory \cite{trugman}, $\nu=4/3$, and $\nu \sim 2.3$, occuring
as the correlation length of the disordered potential vanishes.
A physical interpretation of this intermediate region would be that
the neglected couplings, containing
derivatives, are not expected to contribute when the system is characterized
by disordered potentials correlated in very large scales. 

However, even if $\nu=2$ as the true asymptotic exponent seems
to be a remote possibility, we cannot rule out some conjectures
originated from the fact that the sizes used in numerical computations are
really small, and the range of energies limited by the requirement of
numerical precision. At least, we can propose two possibilities, both based
essentially
on the idea of crossover: 1) there may be a crossover from $\nu \sim 2.3$ to
$\nu=2$ as the energy approaches the center of a broadened Landau level; 2) as
may be inferred from section III, finite size effects break a $O(2)^N$ symmetry
between the $2^N$ fields introduced after $N$ steps of the renormalization
group iteration. It is possible, then, to have a modification of the critical
exponents, reflecting the existence of symmetries which are exact only in the
thermodynamic limit. In this situation, one would expect a crossover to $\nu=2$
, at fixed energy, as the thermodynamic limit is reached. A related observation
is that in conventional Anderson localization, a crossover may be induced
through the breaking of spin and time-reversal symmetries \cite{wegner}.

It would be interesting to apply the present formalism to other close problems
, as, for instance, the analysis of alternative definitions of disordered
potentials and the
computation of conductivity properties at the metal-insulator transition.

\acknowledgments
I would like to thank R. Bhatt, A. Ludwig and M. Moriconi for interesting
discussions. Also, I would like to thank a referee for calling my atention
to ref. \cite{jan}.
This work was supported by CNPq(Brazil).

\appendix
\section{derivation of (3.28)}
In terms of the generating function of Hermite polynomials,
$f(w)=\exp(-w^2+2ws)$, we have, according to (\ref{rg19}),
\bea
&&g_n(2^N(k_1-k_2),2^N(k_3-k_2))= {{\partial^{n}} \over {\partial w_1^n}}
{{\partial^{n}} \over {\partial w_2^n}}
{{\partial^{n}} \over {\partial w_3^n}}
{{\partial^{n}} \over {\partial w_4^n}}
\int dy 
\exp \left[ -(w_1^2+w_2^2+w_3^2+w_4^2) \right. \nonumber \\
&&\left. +2(w_1s_1+w_2s_2+w_3s_3+w_4s_4)
-(s_1^2+s_2^2+s_3^2+s_4^2)/2 \right] \ , \
\label{b1}
\eea
where
\bea
&&s_1=(y-\ell^22^Nk_1)/ \ell \nonumber \\
&&s_2=(y-\ell^22^Nk_2)/ \ell \nonumber \\
&&s_3=(y-\ell^22^Nk_3)/ \ell \nonumber \\
&&s_4=(y-\ell^22^N(k_1-k_2+k_3))/ \ell \label{b2} 
\eea
and the derivatives are taken at $w_1,w_2,w_3,w_4=0$. The integration over $y$
in (\ref{b1}) yields
\bea
&&g_n=\sqrt{{ \pi \over 2}} \ell \exp \left\{ -2^{2N}{ {\ell^2} \over 2}
\left[ (k_1-k_2)^2+(k_3-k_2)^2 \right] \right\}
{{\partial^{n}} \over {\partial w_1^n}}
{{\partial^{n}} \over {\partial w_2^n}}
{{\partial^{n}} \over {\partial w_3^n}}
{{\partial^{n}} \over {\partial w_4^n}}  \nonumber \\
&&\exp [ -(w_1^2+w_2^2+w_3^2+w_4^2)+{1 \over 2}
(w_1+w_2+w_3+w_4)^2+ 2^N \ell (k_1+k_3)(w_1+w_2+w_3+w_4)  \nonumber \\
&&-2^{N+1} \ell (w_1k_1+w_2k_2+w_3k_3+w_4(k_1-k_2+k_3)) ] 
\ . \ \label{b3}
\eea
Using, then (in the sense of distribution theory) that
$\exp(-ax^2) \sim a^{-1/2} \delta (x)$ when $a \rightarrow
\infty$, we get, from (\ref{b3}),
\be
g_n = {{C_n} \over {2^{2N}}} \delta (k_1-
k_2) \delta (k_3-k_2)
\ , \  \label{b4}
\ee
as the leading term when $N \rightarrow \infty$. The coefficient $C_n$ is
writen as
\bea
&&C_n \sim {{\partial^{n}} \over {\partial w_1^n}}
{{\partial^{n}} \over {\partial w_2^n}}
{{\partial^{n}} \over {\partial w_3^n}}
{{\partial^{n}} \over {\partial w_4^n}} 
\exp [ -(w_1^2+w_2^2+w_3^2+w_4^2)+{1 \over 2}
(w_1+w_2+w_3+w_4)^2] \nonumber \\
&&= \int ds \left( H_n(s) \exp(-{1 \over 2} s^2) \right)^4  \ . \ \label{b5}
\eea


\begin{references}
\bibitem{kep} K. Von Klitzing, G. Ebert and M. Pepper, Phys. Rev. Lett. 
{\bf 45}, 494 (1980).
\bibitem{aalr} E. Abrahams, P. W. Anderson, D. C. Licciardello, and
T. V. Ramakrishnan, Phys. Rev. Lett. {\bf 42}, 673 (1979).
\bibitem{hikami} S. Hikami, Prog. Theor. Phys. {\bf 76}, 1210 (1986).
\bibitem{jan} M. Jansen, Int. J. Mod. Phys. B {\bf 8}, 943 (1994).
\bibitem{milsok} G. V. Mil'nikov and I. M. Sokolov, Pis'ma Zh. Eksp. Teor.
Fiz. {\bf 48}, 494 (1988) [JETP Lett. {\bf 48}, 536 (1988)].
\bibitem{gurv} S. Gurvitz, {\it Resonant Scattering on Impurities in the
Integer Quantum Hall Effect}, TRI-PP-94-33, cond-mat@babbage.sissa.it
9406022.
\bibitem{aa} H. Aoki and T. Ando, Phys. Rev. Lett. {\bf 54}, 831 (1985).
\bibitem{chaco} J. T. Chalker and P. D. Coddington, J. Phys. C {\bf  21},
2665 (1988).
\bibitem{huck} B. Huckstein and B. Kramer, Phys. Rev. Lett. {\bf 64},
1437 (1990).
\bibitem{huo} Y. Huo and R. N. Bhatt, Phys. Rev. Lett. {\bf 68},
1375 (1992).
\bibitem{kiv} D.-H. Lee, Z. Wang and S. Kivelson, Phys. Rev. Lett. {\bf 70},
4130 (1993).
\bibitem{liu} D. Liu and S. Das Sarma, Phys. Rev. B {\bf 49}, 2677 (1994).
\bibitem{koch} S. Koch, R. J. Haug, K. Von Klitzing, and K. Ploog,
Phys. Rev. B {\bf 43}, 6828 (1991).
\bibitem{kadanoff} L. P. Kadanoff, Physics(N.Y.) {\bf 2}, 263 (1966).
\bibitem{trugman} S. A. Trugman, Phys. Rev. B {\bf 27}, 7539 (1983).
\bibitem{wegner} F. J. Wegner, Nucl. Phys. {\bf B270}, 1 (1986).
\end{references}
\end{document}